\documentclass[useAMS,usenatbib]{mn2e}
\usepackage{graphicx}

\title[Searching double components]{Searching double-peaked emission line profiles in the spectra of galaxies through the symmetry of the cross-correlation function}

\author[Garc\'{\i}a-Lorenzo et al.]{B. Garc\'{\i}a-Lorenzo$^{1,2}$\thanks{E-mail:
bgarcia@iac.es}\\
1 Instituto de Astrof\'{\i}sica de Canarias, C/Via Lactea S/N, 38305-La Laguna, Tenerife, Spain \\
2 Dept. Astrof\'{\i}sica, Universidad de La Laguna, C/Astrof\'{\i}sico Francisco S\'anchez, E-38205 Tenerife, Spain \\
}

\begin{document}

\date{Accepted ..... Received .....; in original form .....}

\pagerange{\pageref{firstpage}--\pageref{lastpage}} \pubyear{2012}

\maketitle

\label{firstpage}

\begin{abstract} 

The presence of double-peaked/multicomponent emission line profiles in spectra of galaxies is commonly done by visually inspection. However, the identification of complex emission line profiles by eye is unapproachable for large databases such as the Sloan Digital Sky Survey (SDSS) or the integral field spectroscopy surveys of galaxies (e.g. CALIFA or MaNGA). We describe a quick method involving the cross-correlation technique for detecting the presence of complex (double-peaked or multiple components) profiles in the spectra of galaxies, deriving simultaneously a first estimation of the velocity dispersions and radial velocities of the dominant gaseous component. We illustrate the proposed procedure with the well-known complex [OIII]$\lambda\lambda4959,5007$ profiles of the central region of NGC~1068. 

\end{abstract}

\section{Introduction}

Early spectroscopic studies of galaxies showed that profiles of emission lines could present asymmetries, shoulders, or double peaks \citep[e.g.]{heckman81,pelat80,glaspey76,seyfert43}. These features come from the light of several gaseous systems with different kinematics that is integrated along the observer's light of sight. They have been interpreted as due to rotating gaseous disks, outflows/inflows or dual active galactic nuclei \citep[e.g.]{shen11,greene05,zhou04,arribas96}.

Systematic search of double-peaked emission line profiles in large spectra databases (e.g. Sloan Digital Sky Survey, SDSS \citep{york00}) have been performed using different selection criteria, but the final confirmation is usually done by visual inspection \citep{ge12,pilyugin12,smith10,liu10}. In this work, we present an automatic procedure to detect multi-component emission line profiles in large databases of spectra of galaxies based on the symmetry of the cross-correlation function. 

The cross-correlation (C-C hereafter) technique has been extensively used in astronomy to infer radial velocities \citep{anglada12,westfall11,allende07,fromerth00,gunn96,storm92,dalle91,tonry1979}. The C-C technique globally compares a problem spectrum with a reference spectrum or template. The C-C of two spectra analyses the similarity between one spectrum and a wavelength (or velocity) shifted version of the other, as a function of this wavelength (or velocity) shift. The C-C function only contains the frequencies that are common to both spectra. Therefore, the C-C function provides a clear indication of the shift at which the two spectra are most similar and also a quantitative measure of that similarity. As long as the largest peak of the C-C function is symmetric, it might be used to derive the shift of the problem spectrum as well as its velocity dispersion in conjunction with the width of the template \citep{tonry1979}. When the two spectra are the same, the C-C function becomes the auto-correlation.

We propose a methodology for searching multi-component/double-peaked emission line
profiles in the spectra of galaxies based on the deviation from symmetry of the peak
 of the C-C function. Details on the C-C technique can be found in \cite{furenlid90}, \cite{tonry1979} and references therein. The C-C algorithm might be used in the search
 for binary active galactic nuclei in large spectra databases (e.g. SDSS). It may be also used to locate spaxels with spectra showing
 multicomponent emission line profiles in Integral Field Spectroscopic (IFS)
 surveys of galaxies (e.g. the on-going Calar Alto Legacy Integral Field Area Survey, CALIFA \citep{sanchez12} or The Mapping Nearby Galaxies at APO, MANGA (http://dunlap.utoronto.ca/research/surveys/)).

\section{Cross-correlation technique for searching double-peaked line profiles}

\subsection{Cross-correlation function shape traced by bisectors}

\begin{figure*}
\centering
\includegraphics[width=12.25cm]{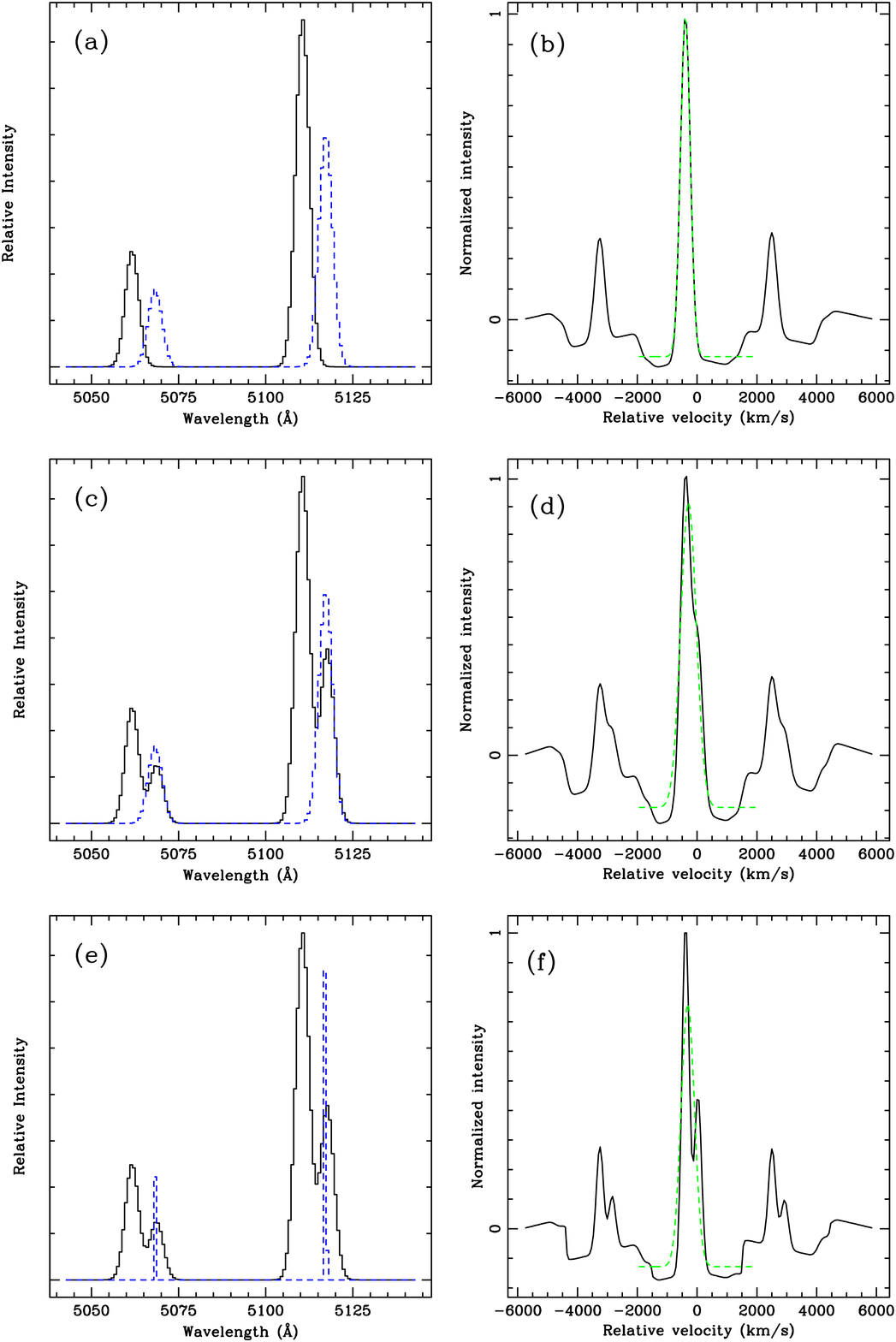}
 \caption{Example of the cross-correlation technique applied in the [OIII]$\lambda\lambda4959,5007$ spectral range, assuming Gaussian emission line profiles and a redshift of about 6610 km/s. Spectral resolution for this model is 1650. (a) Reference spectrum (blue-dashed lines) shifted at the assumed redshift, and problem spectrum (black line) at redshift-400 km/s. Velocity dispersions for problem and template spectrum are 115 km/s. (b) The cross-correlation function of the spectra in (a). The green-dashed lines correspond to the Gaussian profile fit to the peak of the C-C function. The Gaussian centroid (at -390 km/s) and full-width-half-maximum (384.17 km/s) provides the velocity shift between problem and template and the estimation of the problem velocity dispersion, respectively. (c) Problem spectrum (black line) showing double components. The brightest component is described by the problem spectrum (V$_{sys}$=6210 km/s and $\sigma$=115 km/s) in (a), and secondary component is at V$_{sys}$=6710 km/s with a $\sigma$=115 km/s. Both components have been selected to be clearly resolved. Blue-dashed line is the template (the same template than in (a)). (d) The C-C function of the problem and template spectra in (c). The C-C peak clearly has an asymmetric shape although a double-peak is not resolved due to the degradation of the spectral resolution in the calculation of the C-C function. The Gaussian fit to the asymmetric C-C peak provides a velocity shift between template and problem spectra of -289 km/s, and the estimation of the velocity dispersion for the problem is 235 km/s. (e) The velocity dispersion of the template (blue line) is $\sim8$ km/s, negligible compared to the velocity dispersion of the emission lines ($\sigma$=115 km/s) in the problem spectrum (black line). (f) The C-C function of the problem and template spectra in (e). The asymmetric C-C peak profile is now resolved in two peaks. Gaussian fit to the peak of the C-C function provides and estimation of redshift and velocity dispersion of -311 km/s and 225 km/s, respectively, for the problem spectrum. }
\label{model1}
\end{figure*}

The procedure developed in this work is based on the symmetry of the C-C function nearby its main peak. The estimation of the shift and velocity dispersion of the spectra of galaxies might be obtained by fitting a smooth symmetric function to the peak of the C-C function \citep{tonry1979}. However, C-C is not an even function and therefore, this approach is only valid when the shapes of the spectral features in the problem spectrum are similar to those in the reference spectrum. Emission lines in the spectra of galaxies are commonly assumed to be fitted by Gaussian profiles. If single Gaussian profiles are assumed for emission lines in the template and problem spectra, the velocity shift and velocity dispersion of the problem spectrum can be derived by fitting a single Gaussian to the peak of the C-C function (see Fig. \ref{model1}a,b). However, if the emission line profiles in the reference and in the problem spectrum are significantly different, the peak of the C-C function will be asymmetric (see Fig. \ref{model1}c,d). 

The reference spectrum must have a large signal-to-noise ratio and a spectral resolution similar to or better than the problem spectrum. As the C-C of two spectra is similar to their convolution, the spectral resolution of the C-C function might be significantly reduced during the C-C operations. This results in a significant reduction in the capability of detecting double-components (see Fig. \ref{model1}d). However, the template can be selected to have a negligible velocity dispersion compared to the problem spectrum. Indeed, a variation of the C-C technique using delta functions has been successfully used for detection of binary stars \citep{furenlid90}. In this case, the C-C spectral resolution, and hence the capability of detecting double-components, strongly depends on the problem spectrum spectral resolution. Moreover, the shape of the C-C peak will correspond to the average shape of the line profiles in the problem spectrum (figures \ref{model1}e and \ref{model1}f). Therefore, the presence of double or multiple components in emission line profiles is reduced to study the shape of the C-C peak profile. 

\begin{figure*}
\centering
\includegraphics[width=\textwidth]{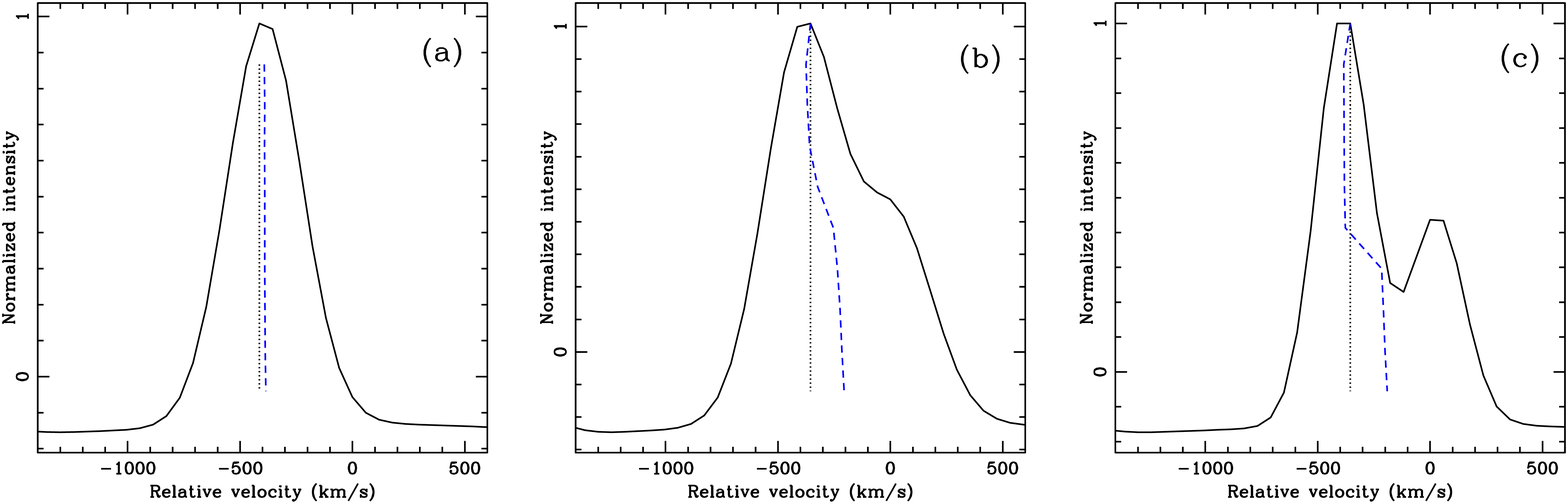}
 \caption{Bisectors (blue dashed-lines) of the C-C peak functions in Fig. \ref{model1}a. Black dotted line corresponds to the peak velocity dividing the profile in two parts. When the emission lines profiles in the template and problem spectrum are similar (Gaussian profiles are assumed) black dotted line divides the C-C peak function profile in two symmetric parts. In this case, bisector (blue dashed-lines) and peak velocity extension to different intensities (black dotted-lines) are similar (see (a)). When asymmetries due to double components are presented in the problem spectrum, the bisector of the C-C peak function is twisted and differs from the velocity at the peak (see (b) and (c)). The difference between the velocity at the peak of the C-C function and the bisector at a determined intensity level indicates the presence of asymmetries (double or multiple components) in the problem spectrum profile. The sign of this difference indicates if these asymmetries are blue or red-shifted respect to the dominant component (peak of the C-C function).}
\label{model2}
\end{figure*}

The C-C peak profile can be characterized by tracing its bisector. Bisector shapes are commonly used in the analysis of the mechanisms that cause asymmetries and variations in stellar spectra \citep[e.g.]{ba11}. The bisector of a symmetric profile should remain at constant wavelength (or velocity) for all parts of the profile, dividing it into two equal parts; any existing asymmetry between the base and the peak of the line will remain reflected in the shape of the bisector. Bisectors for any profile can be constructed by connecting the midpoints of horizontal line segments spanning the width of the profile at a number of intensity positions inside the profile. The comparison of the C-C peak profile bisector (V$_b$) with the C-C peak velocity (V$_p$, the remained value as if it were symmetric) gives a reference value to identify the presence of various gaseous components in the profiles. This procedure does not infer the actual number of gaseous components forming the observed profile, but an evidence of the presence of various gaseous components (at least two). The sign of V$_b$-V$_p$ also gives information about the equivalent wavelength (velocity) of the double/multiple components relative to the dominant component (wavelength or velocity at the C-C peak): if the sign is positive, the secondary (or equivalent) component is red-shifted respect to the dominant component. A positive detection of double/multiple components in a galaxy spectrum will depend on the signal-to-noise of the C-C function and hence on the signal-to-noise of the problem spectrum. Figure \ref{model2} shows the bisectors of the C-C peak functions for the examples in figure \ref{model1}.
Tracing the bisector on a single emission line profile would also provide evidences of the presence of double/multiple components forming that profile. However, any observational or instrumental signature (e.g. cosmic-ray) not properly removed during the data reduction process and affecting the selected single emission line could result in an asymmetric line bisector and hence in an spurious double/multiple component detection. Including several emission lines in the selected cross-correlation spectral range will smoothed any undesirable feature affecting a single-line since the shape of C-C peak function will correspond to their average shape. Moreover, as the C-C function also provides a quantitative measure of the similarity between the problem and template, other advantage of tracing the bisector on the C-C peak function instead of on single emission line profiles is that any shape of the line profiles can be selected to generate the template. Here, we have assumed gaussian profiles for the emission lines to create the template but any other profile shape could be assumed (e.g. Lorentz profiles). 

\subsection{Implementation of the procedure}

The C-C technique uses the information contained in all the lines of the selected spectral range for calculating the C-C function. The resulting shape of the C-C peak corresponds to an average shape of the line profiles in the problem spectrum. 

In order to calculate the C-C function, it is necessary to prepare the problem spectrum following these steps:
\begin{itemize}
\item[1.] The ends of the C-C spectral range are selected at the continuum to avoid discontinuities at the edges.
\item[2.] The data are divided by the continuum to remove any residual curvature in the continuum.
\item[3.] In order to minimize edge effects, the continuum is subtracted (just subtract the unity when step 2 was performed).
\end{itemize}

When a stellar fit had been carried out previously and properly subtracted from the original spectra, steps 2 and 3 can be stripped.

The reference spectrum is generated including as many delta functions as single 
emission lines are expected in the spectral range selected to apply the C-C
 technique. The wavelength (or velocity) position of these delta functions 
corresponds to the center of the emission lines.  Different weights (in flux) can be given
 to the delta functions to account for fix intensity ratio between
 emission lines according to atomic parameters. Noise contribution is not desirable to be included in the template to avoid degradation of the signal-to-noise when the C-C function is computed.
 Both template and problem spectra are normalized to the maximum of the brightest emission line in
 the selected wavelength. When the reference has been generated and the problem spectrum has
 been rectified, the cross-correlation function can be computed. The maximum of
 the cross-correlation function is located and a Gaussian is fitted to
 this peak to obtain a better precision in the determination of the velocity
 shift between template and problem spectra (V$_p$) than the pixel size/resolution element. The bisector of the C-C
 peak function can be traced by calculating the midpoints of horizontal lines segments slicing 
the C-C peak profile at a number of intensities (e.g. twenty intensity positions, from 100\% to
 5\% in steps of 5\%). In order to detect asymmetries, we can compute the difference between V$_p$
 and the bisector at different intensities (e.g. V$_b$(i), for i=1,N, being N the total number of intensity positions). We will have as many V$_p$-V$_b$ as 
intensities have been selected to trace the bisector. If $\Delta$V denotes the size of the resolution element, then:
\begin{itemize}
\item[-] A blue asymmetry is detected when V$_b$-V$_p$ $< -\Delta$V
\item[-] A red asymmetry is detected when V$_b$-V$_p$ $> \Delta$V
\end{itemize}

For the implementation of the procedure, we calculate the number of $|$V$_b$-V$_p$(i)$|$ values (for i=1,N, being N the total number of intensity positions) larger than $\Delta$V (N$_{asym}$) and also the number of V$_b$-V$_p$(i) smaller than -$\Delta$V (N$_{blue}$) and larger than $\Delta$V (N$_{red}$), considering that:

\begin{itemize}
\item When N$_{asym}$=N$_{blue}$ a pure blue asymmetry in the profile is detected, and at least two components are present.
\item When N$_{asym}$=N$_{red}$ a pure red asymmetry in the profile is detected, and at least two components are present.
\item When N$_{asym} >$ N$_{red}$ or/and N$_{asym} >$ N$_{blue}$ then the profile presents blue and red asymmetries depending on the intensity level (multi-asymmetries hereafter), indicating the presence of multiple gaseous components forming the observed emission line profiles.
\end{itemize}

Obviously, the level of noise in the problem spectrum affects the detection of asymmetries. In this sense, a signal-to-noise threshold is defined to reduce the number of false positive detections of double/multiple components on emission line profiles in the spectra of galaxies. Removing bisector lower levels might also help to avoid false positive detections due to poor signal-to-noise of the spectra. In practice, we consider a positive asymmetry detection (blue, red, or multi-asymmetries) when N$_{asym}$ is equal or larger than 2, that is, at least two intensity levels tracing the bisector of the C-C peak function must satisfied that $|$V$_b$-V$_p$(i)$|$ $>\Delta$V.  

\subsection{Application to integral field spectroscopy}

In this section, we present some examples to illustrate the results from the proposed procedure for searching double or multiple gaseous components in the spectra of galaxies applied on IFS data.

\subsubsection{The data}

\begin{figure*}
\centering
\includegraphics[width=\textwidth]{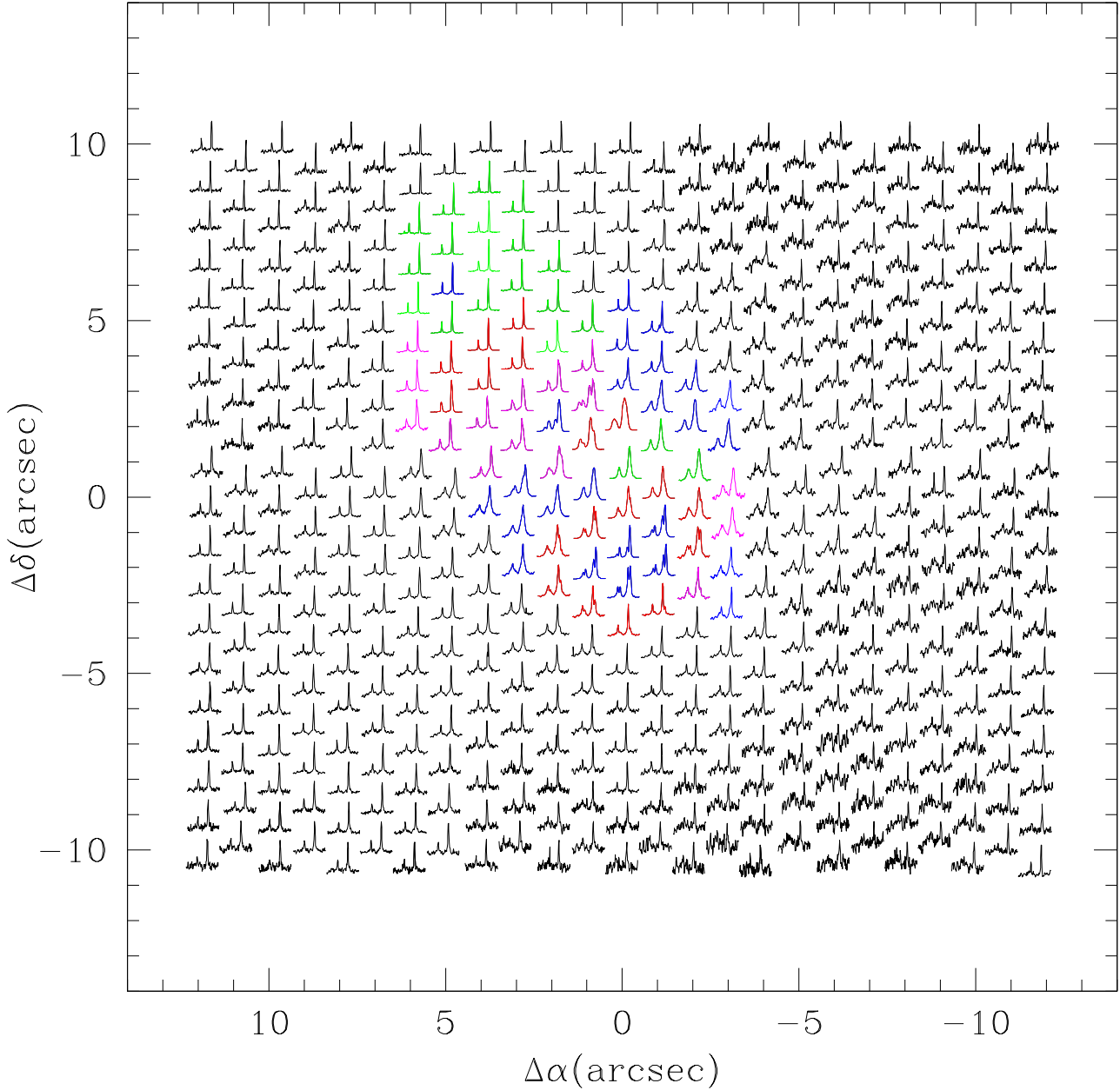}
 \caption{Spectra diagram of the [OIII]$\lambda\lambda4959,5007$ emission lines at 463 positions on the central region of NGC~1068 \citep{garcia99}. The spectra at each location are auto scaled to better show the profile shape. Emission lines nearer to the optical nucleus (taken to be the origin) are brighter than those farther out. The plotted spectral range is 4925-5075 \AA , the same selected for the application of the cross-correlation technique. Different colors indicate the asymmetry detected in those profiles with a signal-to-noise larger than 200 (red: red-asymmetry; blue: blue-asymmetry; magenta: multi-asymmetries; green: symmetric profiles).}
\label{oiii_profiles}
\end{figure*}

We have used the IFS data of the central 24$\mathstrut{^{{\prime}{\prime}}}\times$20$\mathstrut{^{{\prime}{\prime}}}$ of the Seyfert 2 galaxy NGC 1068 presented in \cite{garcia99}. NGC~1068 is the nearest and brightest example of a barred galaxy with an active galactic nucleus. The presence of several kinematically distinct gaseous components in the central region of NGC~1068 is evident from the large amount of emission-line profiles obtained using different instruments and techniques \citep{ozaki09, emsellem06, gerssen06, ishigaki04, groves04, cecil02, garcia99, arribas96, cecil90, pelat80}. Careful visual examination (see Fig. \ref{oiii_profiles}) and description of the spectra \citep{emsellem06}[e.g] revealed the presence of a minimum of three different gaseous systems in NGC~1068. When studying IFS data, the visual inspection of the spectra is a tedious task but accessible for examining a several hundred of spectra. This is not the case for the large amount of spectra provided by IFS surveys (e.g. CALIFA or MaNGA surveys) of galaxies.

Figure \ref{oiii_profiles} shows the [OIII]$\lambda\lambda4959,5007$ line profiles at each observed position of NGC~1068 as a spectra diagram \citep{garcia99}. These profiles offer a large variety of examples to test the developed algorithm.

\subsubsection{Results for NGC~1068}

Applying the proposed procedure of studying the symmetry of the C-C peak function explained in section \S2 to the spectra from the central region of NGC~1068, we can point out the following:

\begin{itemize}
\item Only 12\%  of the emission lines profiles in the central 24$\mathstrut{^{{\prime}{\prime}}}\times$20$\mathstrut{^{{\prime}{\prime}}}$ of the Seyfert 2 galaxy NGC 1068 (Fig. \ref{oiii_profiles}) seems to present symmetric profiles. The percentage of symmetric [OIII]$\lambda\lambda4959,5007$ profiles is 24.5\% for those spectra with a signal-to-noise larger than 200.
\item 6 \% of the total number of spectra presents pure red-asymmetries. Red asymmetries are detected in the 26\% of spectra with a signal-to-noise larger than 200. 
\item 51 \% of the total number of spectra presents pure blue-asymmetries. This percentage is 31.5\% for those spectra with a signal-to-noise ratio $ > 200$. 
\item 31 \% of the spectra in the central region of NGC~1068 present asymmetries, to the blue or red depending on the bisector intensity level. In the case of spectra with a signal-to-noise larger than 200, the percentage of multi-asymmetric emission line profiles is 18\%.
\end{itemize}

\begin{figure*}
\centering
\includegraphics[width=16cm]{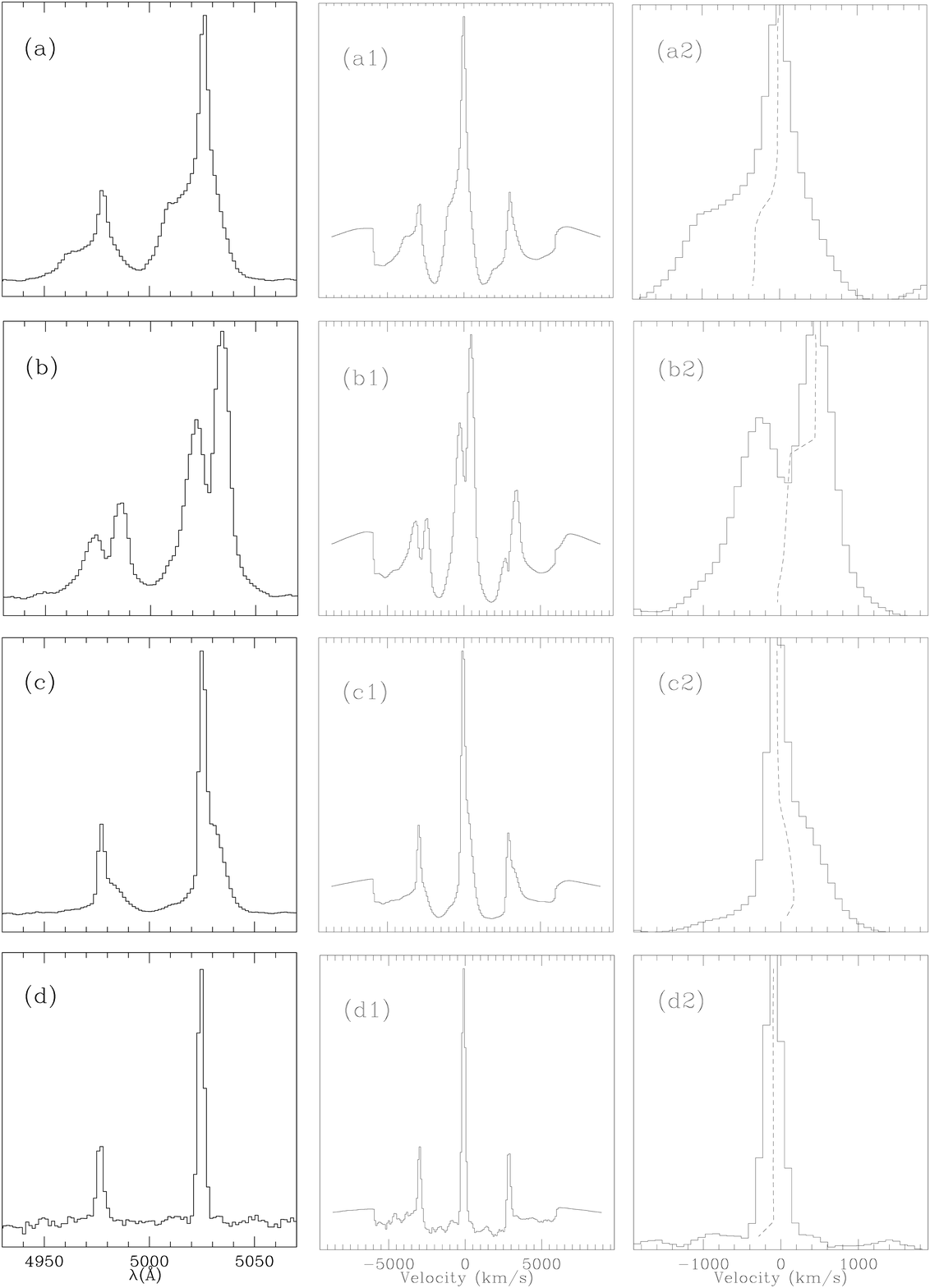}
 \caption{(Left-column) Examples of emission line profiles of NGC~1068. Spectra are located at positions (a) [+0.02,+3.49]; (b) [-1.02,-0.68] arcsec; (c) [+2.97,+4.08] arcsec; and (d) [-8.86,6.24] arcsec from the optical nucleus. (Central-column) (a1),(b1),(c1), and (d1) The cross-correlation functions of the spectra in the left-column ((a),(b),(c), and (d)) using two delta function as template spectrum in the spectral range including the [OIII]$\lambda\lambda4959,5007$ emission lines (4925-5075 \AA ). (Righ-column) The cross-correlation peak function (zoom of function in central column). Dashed line corresponds to the traced bisector.  }
\label{examples}
\end{figure*}

Examples of spectra (the same spectra than in Fig. 7 in Garc\'{\i}a-Lorenzo et al. (1999)) displaying asymmetries and the bisectors shape of the C-C peak function are shown in Fig. \ref{examples}. The [OIII]$\lambda\lambda4959,5007$ profiles in Fig. \ref{examples}(a) show a blue shoulder (corresponding to component 4b in Garc\'{\i}a-Lorenzo et al. (1999) and the additional component in Emsellem et al. (2006)). The shape of the C-C function for this spectrum (Fig. \ref{examples}(a1) and (a2)) also presents a clear blue asymmetry that is traced by the shape of its bisector. The peak of the C-C function indicates a radial velocity of 1145 km/s for the dominant component in this profile, which is in agreement to the systemic velocity, 1144 km/s \citep{emsellem06}. The difference in velocity between the bisector at a 10\% peak intensity level and the velocity at the peak of the C-C function (V$_b$(10\%)-V$_p$) is -302 km/s, indicating a blue asymmetry. The [OIII]$\lambda\lambda4959,5007$ emission lines in Fig. \ref{examples}(b) presents double-peaked [OIII] profiles that it is also reproduced in the C-C peak function. The radial velocity of the dominant component in this profile (velocity at the peak of the C-C function) is 1618 km/s (474 km/s larger than the systemic velocity), which was identified as component 4r in Garc\'{\i}a-Lorenzo et al. (1999) and as the additional component in Emsellem et al (2006). The bisector of the C-C peak function clearly indicates the presence of a secondary bluer component in this profile. The difference V$_b$(10\%)-V$_p$ is -485 km/s. This secondary component was identified as component 1+3 in Garc\'{\i}a-Lorenzo et al. (1999) and as the narrow component in Emsellem et al (2006). Emission lines in Fig. \ref{examples}(c) show a peaked profile with a clear red shoulder that is reproduced by the shape of the C-C peak function bisector. In this case, V$_p$=1124.52 km/s (component 1/narrow in Garc\'{\i}a-Lorenzo et al. (1999)/Emsellem et al. (2006), respectively), while V$_b$(10\%)-V$_p$= 209 km/s, indicating a red component identified as component 3 in Garc\'{\i}a-Lorenzo et al. (1999) and broad component in Emsellem et al (2006). For this profile, the lower level of the bisector (Fig. \ref{examples}(c2)), turns to the blue, V$_b$(5\%)-V$_b$(10\%) = -104 km/s, although still indicating a global red asymmetry (V$_b$(5\%)-V$_p$ = 105 km/s). This twist in the bisector indicates the presence of a faint additional component, that was identified as component 2 in Garc\'{\i}a-Lorenzo et al. (1999). The visual inspection of emission line profiles in Fig. \ref{examples}(d) showed single profiles with not signs of double-components. The C-C function is symmetric respect to its peak, being V$_p$ = 1071 km/s. Only at a 5\% level of intensity of the C-C peak function, a blue asymmetry is detected by the proposed procedure (with a V$_b$(5\%)-V$_p$ = -188 km/s), while for the rest of the levels, the difference V$_b$(i)-V$_p$ (for i=10\%,100\% in steps of 5\%) is always smaller than 10 km/s. This false positive asymmetry detection is due to the low signal-to-noise at the level in which the asymmetry has been detected. In practice, this case will be labeled as symmetric in the implemented algorithm due to the condition of having at least two intensity levels with positive detection for asymmetries. Figure \ref{oiii_profiles} also indicates the identified asymmetries for the observed spectra with a signal-to-noise ratio larger than 200 in the central region of NGC~1068.

The proposed procedure only indicates those spectra showing asymmetries due to double or multiple components, but the deprojection of the different gaseous systems needs any other procedure such as Gaussian fit \citep{arribas96}[e.g.].

\section{Conclusions}

The identification of double/multiple component emission line profiles in spectra of galaxies is the first step for the sample selection of many research topics (e.g. binary black holes, ionized gas kinematic deprojection). This paper deals with a quick estimation of velocity dispersions, radial velocities of the dominant component and the detection of the presence of double or multiple components among a large set of spectra of a galaxy or galaxies using the cross-correlation technique and the shape of the C-C peak function. The proposed procedure allows processing a large amount of problem spectra in a short period of time (in the order of minutes) using the peak, full-width-half-maximum and symmetry of the C-C peak function. 

\section*{Acknowledgments}

The author thanks A. Eff-Darwich, and A. Bongiovanni for their help and useful discussions. This work was partially funded by the Instituto de Astrof\'{\i}sica de Canarias, by the Spanish Ministerio de Econom\'{\i}a y Competitividad (MINECO; grant AYA2009-12903) and by the Spanish Agencia Canaria de Investigaci\'on, Innovaci\'on y Sociedad de la Informaci\'on (proID20100121).


\begin{thebibliography}{99}

\bibitem[\protect\citeauthoryear{Allende Prieto}{2007}]{allende07} Allende Prieto, C.\ 2007, AJ, 134, 1843 


\bibitem[\protect\citeauthoryear{Anglada-Escud{\'e} \& Butler}{2012}]{anglada12} Anglada-Escud{\'e}, G., \& Butler, R.~P.\ 2012, ApJSS, 200, 15 

\bibitem[\protect\citeauthoryear{Arribas et al.}{ 1996}]{arribas96} Arribas, S., 
Mediavilla, E., \& Garcia-Lorenzo, B.\ 1996, ApJ, 463, 509 

\bibitem[\protect\citeauthoryear{Ba{\c s}t{\"u}rk et al.}{2011}]{ba11} Ba{\c s}t{\"u}rk, {\"O}., Dall, T.~H., Collet, R., Lo Curto, G., \& Selam, S.~O.\ 2011, A\&A, 535, A17 

\bibitem[\protect\citeauthoryear{Cecil et al.}{ 2002}]{cecil02} Cecil, G., Dopita, M.~A., Groves, B., et al.\ 2002, ApJ, 568, 627 

\bibitem[\protect\citeauthoryear{Cecil et al.}{ 1990}]{cecil90} Cecil, G., Bland, J., 
\& Tully, R.~B.\ 1990, ApJ, 355, 70 

\bibitem[\protect\citeauthoryear{Dalle Ore et al.}{ 1991}]{dalle91} Dalle Ore, C., Faber, 
S.~M., Jesus, J., Stoughton, R., \& Burstein, D.\ 1991, ApJ, 366, 38 

\bibitem[\protect\citeauthoryear{Emsellem et al.}{ 2006}]{emsellem06} Emsellem, E., Fathi, 
K., Wozniak, H., et al.\ 2006, MNRAS, 365, 367 

\bibitem[\protect\citeauthoryear{Fromerth \& Melia}{ 2000}]{fromerth00} Fromerth, M.~J., \& Melia, F.\ 2000, ApJ, 533, 172 


\bibitem[\protect\citeauthoryear{Furenlid \& Furenlid}{ 1990}]{furenlid90} Furenlid, I., \& Furenlid, L.\ 1990, PASP, 102, 592 

\bibitem[\protect\citeauthoryear{Garc{\'{\i}}a-Lorenzo et al.}{ 1999}]{garcia99} 
Garc{\'{\i}}a-Lorenzo, B., Mediavilla, E., \& Arribas, S.\ 1999, ApJ, 518, 190 

\bibitem[\protect\citeauthoryear{Ge et al.}{2012}]{ge12} Ge, J.-Q., Hu, C., Wang, 
J.-M., Bai, J.-M., \& Zhang, S.\ 2012, ApJSS, 201, 31 


\bibitem[\protect\citeauthoryear{Gerssen et al.}{ 2006}]{gerssen06} Gerssen, J., Allington-Smith, J., Miller, B.~W., Turner, J.~E.~H., 
\& Walker, A.\ 2006, MNRAS, 365, 29 

\bibitem[\protect\citeauthoryear{Glaspey et al.}{1976}]{glaspey76} Glaspey, J.~W., Eilek, 
J.~A., Fahlman, G.~G., \& Auman, J.~R.\ 1976, ApJ, 203, 335 

\bibitem[Greene \& Ho(2005)]{greene05} Greene, J.~E., \& Ho, L.~C.\ 2005, ApJ, 627, 721 


\bibitem[\protect\citeauthoryear{Groves et al.}{ 2004}]{groves04} Groves, B.~A., Cecil, 
G., Ferruit, P., \& Dopita, M.~A.\ 2004, ApJ, 611, 786 

\bibitem[\protect\citeauthoryear{Gunn et al.}{ 1996}]{gunn96} Gunn, A.~G., Hall, J.~C., Lockwood, G.~W., \& Doyle, J.~G.\ 1996, A\&A, 305, 146 


\bibitem[\protect\citeauthoryear{Heckman et al.}{1981}]{heckman81} Heckman, T.~M., Miley, 
G.~K., van Breugel, W.~J.~M., \& Butcher, H.~R.\ 1981, ApJ, 247, 403 


\bibitem[\protect\citeauthoryear{Ishigaki et al.}{ 2004}]{ishigaki04} Ishigaki, T., Hayashi, 
T., Ohtani, H., et al.\ 2004, PASJ, 56, 723 

\bibitem[\protect\citeauthoryear{Liu et al.}{ 2010}]{liu10} Liu, X., Shen, Y., Strauss, 
M.~A., \& Greene, J.~E.\ 2010, ApJ, 708, 427 

\bibitem[\protect\citeauthoryear{Ozaki}{ 2009}]{ozaki09} Ozaki, S.\ 2009, PASJ, 61, 259 

\bibitem[\protect\citeauthoryear{Pelat \& Alloin}{ 1980}]{pelat80} Pelat, D., \& Alloin, D.\ 1980, A\&A, 81, 172 

\bibitem[\protect\citeauthoryear{Pilyugin et al.}{ 2012}]{pilyugin12} Pilyugin, L.~S., 
Zinchenko, I.~A., Cedr{\'e}s, B., et al.\ 2012, MNRAS, 419, 490 

\bibitem[\protect\citeauthoryear{S{\'a}nchez et al.}{2012}]{sanchez12} S{\'a}nchez, S.~F., Kennicutt, R.~C., Gil de Paz, A., et al.\ 2012, A\&A, 538, A8 

\bibitem[\protect\citeauthoryear{Seyfert}{1943}]{seyfert43} Seyfert, C.~K.\ 1943, ApJ, 97, 28 

\bibitem[Shen et al.(2011)]{shen11} Shen, Y., Liu, X., Greene, 
J.~E., \& Strauss, M.~A.\ 2011, ApJ, 735, 48 

\bibitem[\protect\citeauthoryear{Smith et al.}{ 2010}]{smith10} Smith, K.~L., Shields, 
G.~A., Bonning, E.~W., et al.\ 2010, ApJ, 716, 866 

\bibitem[\protect\citeauthoryear{Storm et al.}{1992}]{storm92} Storm, J., Carney, B.~W., 
Latham, D.~W., Davis, R.~J., \& Laird, J.~B. 1992, PASP, 104, 168 

\bibitem[\protect\citeauthoryear{Tonry \& Davis}{ 1979}]{tonry1979} Tonry, J., \& Davis, M.\ 1979, AJ, 84, 1511 

\bibitem[\protect\citeauthoryear{Veilleux}{1991}]{1991ApJS...75..357V} Veilleux, S.\ 1991, ApJSS, 75, 357 
\bibitem[\protect\citeauthoryear{Westfall et al.}{2011}]{westfall11} Westfall, K.~B., 
Bershady, M.~A., \& Verheijen, M.~A.~W.\ 2011, ApJSS, 193, 21 

\bibitem[\protect\citeauthoryear{York et al.}{2000}]{york00} York, D.~G., Adelman, J., 
Anderson, J.~E., Jr., et al.\ 2000, AJ, 120, 1579 

\bibitem[\protect\citeauthoryear{Zhou et al.}{2004}]{zhou04} Zhou, H., Wang, T., Zhang, 
X., Dong, X., \& Li, C.\ 2004, ApJL, 604, L33 

\end{thebibliography}
\end{document}